\documentclass[12pt]{article}
\usepackage{fullpage}
\usepackage{amsmath}
\usepackage{amssymb}
\usepackage[dvips]{epsfig}

\def\##1{\underline{#1}}
\def\=#1{\underline{\underline{#1}}}

\def\+#1{\underline{\bf #1}}
\def\*#1{\underline{\underline{\bf #1}}}

\def\r#1{(\ref{#1})}
\def\l#1{\label{#1}}
\def\c#1{\cite{#1}}

\def\le{\left(}
\def\ri{\right)}
\def\les{\left[}
\def\ris{\right]}
\def\lec{\left\{}
\def\ric{\right\}}

\def\.{\mbox{ \tiny{$^\bullet$} }}

\def\epso{\epsilon_{\scriptscriptstyle 0}}

\def\muo{\mu_{\scriptscriptstyle 0}}

\def\ko{k_{\scriptscriptstyle 0}}

\def\Eo{\#E_{\, \scriptscriptstyle 0}}
\def\Ho{\#H_{\, \scriptscriptstyle 0}}

\def\ux{\hat{\#u}_x}
\def\uy{\hat{\#u}_y}
\def\uz{\hat{\#u}_z}

\begin{document}

\begin{center}

{\bf {\Large Negative phase velocity of
electromagnetic waves and the cosmological constant
}}

 \vspace{10mm} \large
Tom G. Mackay,\footnote{Corresponding Author. Fax: + 44 131
650 6553; e--mail: T.Mackay@ed.ac.uk.}
Sandi Setiawan\footnote{E--mail: S.Setiawan@ed.ac.uk.}\\
{\em School of Mathematics,
University of Edinburgh, Edinburgh EH9 3JZ, UK}\\
\bigskip
\ Akhlesh  Lakhtakia\footnote{E--mail: akhlesh@psu.edu; also
 affiliated with Department of Physics, Imperial College, London SW7 2 BZ,
UK}\\
 {\em Department of Engineering Science and
Mechanics\\ Pennsylvania State University, University Park, PA
16802--6812, USA}\\

\end{center}

\vspace{4mm}

\normalsize

\begin{abstract}
\noindent
Examining the propagation of electromagnetic plane waves with the wavevector directed in opposition to the
time--averaged Poynting vector in cosmological spacetime with piecewise uniform metric, we show that
such negative--phase--velocity (NPV) propagation is possible in certain de Sitter spacetimes but not in anti--de Sitter spacetimes.
This difference suggests the possibility of an optical/electromagnetic experiment to discern the cosmological constant of a
four--dimensional universe stemming from a five--dimensional brane universe.
\end{abstract}

\noindent {\bf Keywords:}  General theory of relativity, Negative phase
velocity, Poynting vector, cosmological constant

\section{Introduction}

Literature on the phenomenon of negative refraction by certain linear, homogeneous, isotropic, dielectric--magnetic materials
began to record an impressive rate of growth from mid--2000, after the announcement of an experimental
result that could not be explained in any other way \c{SSS}. The underlying reason for negative refraction is that
the time--averaged Poynting vector and the wavevector of a plane wave are oppositely aligned in those materials,
for which reason we call them {\em negative--phase--velocity\/} (NPV) materials \c{LMW}.
The promise of industrial exploitation of negative refraction has midwifed the extension of theoretical studies to bianisotropic
materials \c{LMW,pen2,ML_PRE}. The hope is that nanotechnological routes will allow negative refraction in the optical regime,
which would be a truly remarkable accomplishment.

The classical electromagnetic vacuum cannot support NPV plane waves, as it appears the same to all observers
moving at constant relative velocities \c{ML04a}. However, gravitational fields due to nearby massive objects would certainly distort
electromagnetic propagation, which points towards the possibility of gravitationally assisted NPV propagation
in vacuum. Investigation of a special case showed that this possibility cannot be discounted in spacetime manifolds of limited
extent wherein the metric can be assumed to be approximately uniform \c{lm1}.

Our objective in this communication is to examine the facilitation
of  NPV propagation of electromagnetic waves by the cosmological
constant. Applying the formalism presented elsewhere
\c{lm1,MLS_NJP}, we consider both de Sitter and anti--de Sitter
spacetimes \c{gh1977,exact}.

\section{Cosmological spacetime}

In a static (i.e., time-independent) cosmological spacetime, the matrix representation of the
metric $g_{\alpha \beta}$ is expressed in Cartesian coordinates
as\footnote{Roman indexes take the values 1, 2 and 3; while Greek indexes take the values 0, 1, 2, and 3.}
 \c{gh1977}
\begin{eqnarray}
\les \, g_{\alpha \beta} \, \ris & = &  \left( \begin{array}{cccc}
m & 0 & 0 & 0\\
\\
0 & - \le 1 + \frac{\displaystyle \Lambda x^2}{\displaystyle 3c^2 m } \ri & -\frac{\displaystyle \Lambda xy}{\displaystyle 3c^2 m } & -\frac{\displaystyle \Lambda xz}{\displaystyle 3c^2 m } \\
\\
0 & -\frac{\displaystyle \Lambda xy}{\displaystyle 3c^2 m } & - \le 1 + \frac{\displaystyle \Lambda y^2}{\displaystyle 3c^2 m } \ri & -\frac{\displaystyle \Lambda yz}{\displaystyle 3c^2 m } \\
\\
0 & -\frac{\displaystyle \Lambda xz}{\displaystyle 3c^2 m } & -\frac{\displaystyle \Lambda yz}{\displaystyle 3c^2 m } & - \le 1 + \frac{\displaystyle \Lambda z^2}{\displaystyle 3c^2 m } \ri
\end{array} \right)
\end{eqnarray}
where
\begin{equation}
m = 1 - \frac{\displaystyle \Lambda \le x^2 + y^2 + z^2 \ri }{\displaystyle 3c^2}\,,
\end{equation}
and $c$ is the speed of light in vacuum in the absence of a
gravitational field.  If the
cosmological constant
 $\Lambda$ is positive (negative), the spacetime is called de Sitter (anti-de Sitter) spacetime \c{exact}.
The apparent singularity  at $m = 0$ arises because of an infelicitous
choice of coordinate system. The metric can actually be extended to a geodesically complete space of constant curvature in other coordinate systems \c{hawkingellis}.

The electromagnetic response of vacuum in curved spacetime may be described by
the constitutive relations of an equivalent, instantaneously responding, medium 
  as per \c{lm1,Skrotskii,Plebanski,SS}
\begin{equation}
\left.
\begin{array}{l}
\#D = \epso \=\gamma \. \#E
\\
\#B= \muo \=\gamma \. \#H
\end{array}
\right\}, \l{CR}
\end{equation}
wherein SI units are implemented.
Here,  $\epso = 8.854\times 10^{-12}$~F~m~$^{-1}$,
$\muo = 4\pi\times 10^{-12}$~H~m$^{-1}$,
and $\=\gamma$ is  the 3$\times$3 dyadic equivalent of the metric $\les \, \gamma_{ab} \, \ris $ with components
\begin{equation}
\gamma_{ab}  = -  \frac{g^{ab}}{g_{00}} .
\end{equation}

The constitutive relations \r{CR} provide a global description of
cosmological spacetime. Let us partition the global spacetime into
 neighbourhoods \c{MLS_NJP,LMS05}. We focus our attention upon a neighbourhood
${\cal R}$ of the arbitrary location $\le \tilde{x}, \tilde{y},
\tilde{z} \ri$ wherein the nonuniform metric $\gamma_{ab}$ may  be
approximated by the uniform metric $\tilde{\gamma}_{ab}$.
 Thus, we have the uniform 3$\times$3 dyadic
representation
\begin{equation}
\={\tilde\gamma}  \equiv \les \, \tilde{\gamma}_{ab} \, \ris =  \frac{1}{\tilde{m}}
\left( \begin{array}{ccc}
  1-\frac{\displaystyle \Lambda \tilde{x}^2}{\displaystyle 3c^2}
 & -\frac{\displaystyle \Lambda \tilde{x}\tilde{y}}{\displaystyle 3c^2 } & -\frac{\displaystyle
\Lambda \tilde{x}\tilde{z}}{\displaystyle 3c^2 } \\
\\
-\frac{\displaystyle \Lambda \tilde{x}\tilde{y}}{\displaystyle 3c^2  } &  1-\frac{\displaystyle \Lambda \tilde{y}^2}{\displaystyle 3c^2}
 & -\frac{\displaystyle \Lambda \tilde{y}\tilde{z}}{\displaystyle 3c^2  }  \\
\\
-\frac{\displaystyle \Lambda \tilde{x}\tilde{z}}{\displaystyle 3c^2  }
& -\frac{\displaystyle \Lambda \tilde{y}\tilde{z}}{\displaystyle 3c^2 } &  1-\frac{\displaystyle \Lambda \tilde{z}^2 }{\displaystyle 3c^2}
\end{array} \right) \, \l{gamma_m}
\end{equation}
in ${\cal R}$, where the constant
\begin{equation}
\tilde{m} = 1 - \frac{\displaystyle \Lambda \le  \tilde{x}^2 + \tilde{y}^2 + \tilde{z}^2 \ri }{\displaystyle 3c^2}\,.
\end{equation}

\section{Plane waves in ${\cal R}$}
Planewave solutions
\begin{eqnarray}
\#E &=& {\rm Re} \lec\Eo \exp \les i \le  \#k \. \#r - \omega t \ri
\ris\ric\,, \l{pw_e}
\\   \#H &=& {\rm Re} \lec\Ho \exp \les i \le  \#k \. \#r - \omega t \ri
\ris\ric\,,
  \l{pw_h}
\end{eqnarray}
are sought to the source--free Maxwell curl postulates
\begin{eqnarray}
&& \nabla \times \#E + \frac{\partial}{\partial t} \#B = \#0\,, \l{MP_1} \\
&& \nabla \times \#H - \frac{\partial}{\partial t} \#D = \#0\, \l{MP_2}
\end{eqnarray}
in ${\cal R}$. Here, $\#k$ is the wavevector, $\#r$ is the
position vector within the neighbourhood  containing  $\le
\tilde{x}, \tilde{y}, \tilde{z} \ri$, $\omega$ is the angular
frequency, and $t$ denotes the time; whereas $i=\sqrt{-1}$, and
$\Eo$ as well as $\Ho$ are complex--valued amplitudes. By
combining \r{pw_e}--\r{MP_2} we find, after some manipulation,
that
\begin{equation}
\=W \. \Eo = \#0 \,, \l{ev_eq}
\end{equation}
where
\begin{equation}
\=W = \le \ko^2 \mbox{det} \les \, \={\tilde\gamma} \, \ris -
\#k\. \={\tilde\gamma} \. \#k \ri \, \=I + \#k\,\#k \.
\={\tilde\gamma},
\end{equation}
and  $\ko = \omega \sqrt{\epso \muo}$. The corresponding
dispersion relation ${\rm det} \les \, \=W \, \ris = 0$ is
expressible as
\begin{equation}
\ko^2\,  \mbox{det} \les \, \={\tilde\gamma} \, \ris
 \le \ko^2 \,\mbox{det} \les \, \={\tilde\gamma} \, \ris - \#k\. \={\tilde\gamma} \. \#k \ri^2 = 0\,.
\end{equation}
Thus,
 the wavevectors
satisfy the condition
\begin{equation}
\#k\.\={\tilde\gamma}\.\#k = \ko^2 \, \mbox{det} \, \les \, \={\tilde\gamma} \,
\ris\,, \l{disp_cond}
\end{equation}
as long as  $\={\tilde\gamma}$ is nonsingular.

Now we turn to
the eigensolutions of \r{ev_eq}. By virtue of \r{disp_cond}, we have
\begin{equation}
\#k \, \#k \.  \={\tilde\gamma} \. \Eo = \#0 \,; \l{ev_cond}
\end{equation}
hence, $\Eo$ is  orthogonal to $\#k \.
\={\tilde\gamma}$. Without any loss of generality, let us  choose
the wavevector $\#k$ to lie along the $z$ axis, i.e.,
\begin{equation}
\#k = k \uz\, ,
\end{equation}
with
the
unit vector $\uz$ lying along the Cartesian $z$  axis.
Thereby,
\begin{equation} \#k \. \={\tilde\gamma} =
k(\tilde{\gamma}_1 \ux + \tilde{\gamma}_2 \uy +
\tilde{\gamma}_3\uz)\,,
\end{equation}
with
\begin{equation}
\tilde{\gamma}_1 =  - \frac{\displaystyle  \Lambda \tilde{x}\tilde{z}}{ \displaystyle 3 c^2 \tilde{m}} \, ,\quad
\tilde{\gamma}_2 = - \frac{\displaystyle  \Lambda \tilde{y}\tilde{z}}{\displaystyle 3c^2 \tilde{m} } \, ,\quad
\tilde{\gamma}_3 = \frac{\displaystyle \tilde{m}_z}{\displaystyle
\tilde{m}}\,,\quad
\tilde{m}_z = 1-\frac{ \Lambda \tilde{z}^2 }{ 3 c^2}\,,
\end{equation}
and $\ux$ and $\uy$ being unit vectors  lying along the Cartesian $x$ and $y$  axes, respectively.

 Two linearly independent eigenvectors satisfying \r{ev_cond}
are provided as
\begin{eqnarray}
\#e_{\,1} &=& \tilde{\gamma}_2 \ux - \tilde{\gamma}_1 \uy \,, \l{e1} \\
\#e_{\,2} &=& \tilde{\gamma}_1 \tilde{\gamma}_3 \ux
+\tilde{\gamma}_2 \tilde{\gamma}_3 \uy - \le \tilde{\gamma}_1^2 +
\tilde{\gamma}_2^2 \ri \uz \,; \l{e2}
\end{eqnarray}
hence, the general solution is given by
\begin{equation}
\Eo = C_1 \#e_{\,1} + C_2 \#e_{\,2}
\,, \l{E_vec}
\end{equation}
wherein $C_1$ and $C_2$ are arbitrary complex--valued constants.
The  corresponding  expression for $\Ho$ follows  from the Maxwell
postulates as
\begin{equation}
\Ho = \frac{k}{\omega \muo}
 \, \le \, C_1 \,
\tilde{m} \#e_{\,2} - C_2 \tilde{m}_z \#e_{\,1}  \, \ri
 \l{H_vec} \,.
\end{equation}

The wavenumbers emerge straightforwardly  from the dispersion equation \r{disp_cond}.
For $\#k$ aligned with the $z$ axis we obtain the $k$--quadratic expression
\begin{equation}
k^2 \frac{\displaystyle \tilde{m}_z}{\tilde{m}} -
\frac{\displaystyle \ko^2}{\displaystyle \tilde{m}^2} = 0 \,,
\l{quad}
\end{equation}
which yields the wavenumbers
\begin{eqnarray}
k &=&  \pm \ko \, \le \tilde{m}  \tilde{m}_z  \ri^{-1/2}.
 \end{eqnarray}
The requirement that $k \in \mathbb{R}$ imposes the condition
\begin{equation}
\tilde{m}  \tilde{m}_z > 0\,; \l{mm_cond}
\end{equation}
in other words, both $\tilde{m}$ and ${\tilde{m}}_z$ must be of the same signs for real--valued $k$.

\section{NPV Condition}

In order to establish the feasibility of NPV propagation, we turn
to the time--averaged Poynting vector given by
\begin{eqnarray}
\#P &=& \frac{1}{2}\, {\rm Re}\lec \Eo\times \Ho^\ast\ric\,.
\l{P_def}
\end{eqnarray}
The general solution \r{E_vec} and \r{H_vec} delivers
\begin{equation}
\#P = \frac{k}{2\,\omega \muo} \le \, | C_1 |^2 \tilde{m} + | C_2
|^2 \tilde{m}_z \, \ri \, \#e_{\,1} \times \#e_{\,2}\,.
\end{equation}
From \r{e1} and \r{e2} we
have
\begin{equation}
\#e_1 \times \#e_2 = \le  \gamma^2_1 + \gamma^2_2  \ri  \le \gamma_1 \ux +
\gamma_2 \uy +  \gamma_3 \uz  \ri;
\end{equation}
thus
\begin{equation}
\#k \. \le  \#e_1 \times \#e_2 \ri = k
\frac{\tilde{m}_z}{\tilde{m}}  \le \, \frac{ \Lambda \tilde{z}}{3
c^2 \tilde{m}} \, \ri^2 \le \tilde{x}^2 + \tilde{y}^2 \ri
\end{equation}
and
\begin{eqnarray}
\#k \. \#P &=& \frac{1}{2\, \omega \muo} \le \, \frac{k \Lambda
\tilde{z}}{3 c^2 \tilde{m}} \, \ri^2 \le \tilde{x}^2 + \tilde{y}^2
\ri \le \, | C_1 |^2  + | C_2 |^2 \frac{\tilde{m}_z}{\tilde{m}} \,
\ri \tilde{m}_z\,.
\end{eqnarray}

The definition of NPV is that $\#k \. \#P < 0$ \c{ML04a,lm1}. By
exploiting \r{mm_cond}, we see that  $\#k \. \#P < 0$ follows from
$\tilde{m}_z < 0$ ; i.e., NPV propagation arises provided that
\begin{equation}
\Lambda > \frac{3c^2}{\tilde{z}^2}\,.
\l{NPVc}
\end{equation}
According to the NPV condition \r{NPVc}, the anti--de Sitter
spacetime (i.e., $\Lambda < 0$ ) does not support NPV propagation,
unlike the de Sitter spacetime  (i.e., $\Lambda > 0$ ) which
supports NPV propagation as long as the cosmological constant is
sufficiently large. Equivalently, NPV propagation is feasible in
de Sitter spacetime at sufficiently remote locations. We observe
that de Sitter  NPV propagation is supported at locations which
lie outside the event horizon specified by $\le x^2 + y^2 + z^2
\ri = 3 c^2 / \Lambda$.

\section{Applicability of NPV condition}

Let us  comment now on the applicability of the NPV condition
\r{NPVc}. Suppose
   $\delta$ is some representative  linear dimension
of the neighbourhood $\cal R$. The uniform approximation, as
implemented for our planewave analysis, rests upon the assumptions
that $\delta$ is (i) small compared with the curvature of the
global spacetime; and (ii) large compared with electromagnetic
wavelength as given by $2 \pi/k$. To be specific, consider the
Ricci scalar $R$. For de Sitter spacetime we have \c{Peebles}
\begin{equation}
R = \frac{4 \Lambda}{c^2}\,.
\end{equation}
As $R$  provides a measure of the inverse radius of spacetime
curvature squared, the neighbourhood $\cal R$  is such that
\begin{equation}
\frac{2 \pi}{| k |}  \ll \delta \ll \frac{c}{2} \sqrt{\frac{\rho}
{| \Lambda | }}\,, \l{partition}
\end{equation}
where $\rho$ is a proportionality constant. Thus,  we see that the
partition of global spacetime requires
\begin{equation}
| \Lambda | \ll  \frac{c^2 | k |^2 \rho }{16 \pi^2}\,.
\end{equation}
Since the   linear dimensions $\delta$ of the neighbourhood ${\cal
R}$ are chosen independently of the  $\tilde{z}$ coordinate
specifying the location of  ${\cal R}$,  there is no
incompatibility between the conditions imposed on $\Lambda$ by the
NPV inequality \r{NPVc} and the  uniform approximation
inequalities \r{partition}.

 After solving for the planewave propagation
modes in each neighbourhood,  the  neighbourhood solutions may be
stitched together to provide the global solution. The
 piecewise uniform approximation technique adopted here~---~which
  is described in detail elsewhere \c{LMS05}~---~is commonly employed
  in solving differential equations with nonhomogeneous
 coefficients  \c{Hoff}.

\section{Concluding Remarks}

We have theoretically examined the propagation of electromagnetic plane waves with the
time--averaged Poynting vector directed in opposition to the
wavevector in cosmological spacetime with piecewise uniform metric.
Our derivations lead us to conclude that NPV propagation is possible for $\Lambda$ sufficiently large and positive (de Sitter spacetime); but it
 is impossible for negative $\Lambda$, i.e., when the spacetime is of the anti--de Sitter type.
This difference between the two types of spacetimes should be
added to the catalogue of known differences for scalar waves
\c{mk2004}.

Recent string--inspired theories with \emph{large} extra
dimensions, known as brane--world models
 \c{rs1,rs2,add}, suggest
 that gravitational interactions between particles on the brane in uncompactified bulk five-dimensional space
 can exhibit
 the correct
 four--dimensional brane behaviour,  through   finely tuning a relation
 between the bulk cosmological constant and the
 brane tension. In the original Randall--Sundrum model \c{rs2}, the cosmological constant and
the brane tension exactly cancel each other. More recent  models,
which  admit the possibility
 of the cosmological constant and the brane tension  not being in  complete cancellation, yield
  a net cosmological constant on the four--dimensional brane that
 could be either positive or negative \c{bajc}. As a  consequence of our results,
one could envision  an optical/electromagnetic experiment, based
upon the negative refraction associated with NPV propagation
\c{pen2,pen1}, which   may reveal  whether or not a
four--dimensional
  universe~---~deriving from a five--dimensional brane
 universe~---~has a
sufficiently  large positive cosmological constant.

\vspace{10mm}

\noindent{\bf Acknowledgements}
SS acknowledges EPSRC for support under grant GR/S60631/01.


\begin{thebibliography}{99}

\bibitem{SSS}
Shelby R A,  Smith D R and Schultz S 2001
Experimental verification of a negative index of refraction
\emph{Science} {\bf 292} 77--79

\bibitem{LMW}
Lakhtakia A, McCall M W and  Weiglhofer W S 2003
Negative phase--velocity mediums
  \emph{Introduction to Complex Mediums for Optics and
Electromagnetics}  ed  W S Weiglhofer and A Lakhtakia
(Bellingham, WA: SPIE  Press) pp347--363


\bibitem{pen2}
Pendry J B 2004
Negative refraction
\emph{Contemporary Phys.} {\bf 45}, 191--202

\bibitem{ML_PRE}
Mackay TG and  Lakhtakia A 2004
Plane waves with negative phase velocity in Faraday chiral mediums
\emph{Phys. Rev. E} {\bf 69} 026602

\bibitem{ML04a}
Mackay T G and Lakhtakia A 2004 Negative phase velocity in a
uniformly moving, homogeneous, isotropic, dielectric--magnetic
medium \emph{J. Phys. A: Math. Gen.} {\bf 37}  5697--5711


\bibitem{lm1}
Lakhtakia A and  Mackay T G 2004 Towards gravitationally assisted
negative refraction of light by vacuum \emph{J. Phys. A: Math.
Gen.\/} {\bf 37}  L505--L510; corrections: 2004 {\bf 37} 12093

\bibitem{MLS_NJP}
Mackay T G, Lakhtakia A and Setiawan  2005 Gravitation and
electromagnetic wave propagation with negative phase velocity
\emph{New J. Phys.} {\bf 7} 75

\bibitem{gh1977}
Gibbons G W and Hawking S W 1977
Cosmological event horizons, thermodynamics, and particle creation
\emph{Phys. Rev. D} {\bf 15} 2738--2751

\bibitem{exact}
Stephani H, Kramer D, MacCallum M, Hoenselaers C and Herlt E 2003
{\em Exact Solutions of Einstein's Field Equations, 2nd ed\/}
(Cambridge: Cambridge University Press) Chap 8

\bibitem{hawkingellis}
Hawking S W and Ellis G F R 1973 {\em The Large Scale Structure of Space-time\/}
(Cambridge: Cambridge University Press) Chap 5

\bibitem{Skrotskii}
Skrotskii  G V 1957 The influence of gravitation on the propagation of light
\emph{Soviet Phys.--Dokl.\/} {\bf 2}  226--229


\bibitem{Plebanski}
Plebanski J 1960 Electromagnetic waves in gravitational fields
\emph{ Phys. Rev.\/} {\bf 118}, 1396--1408

 \bibitem{SS}
Schleich W  and Scully M O 1984 General relativity and modern optics
 \emph{New Trends in Atomic Physics}
ed  G Grynberg  and R Stora  (Amsterdam, Holland: 
Elsevier Science Publishers) pp995--1124

\bibitem{LMS05}
Lakhtakia A, Mackay T G and Setiawan S 2005  Global and local
perspectives of gravitationally assisted negative--phase--velocity
propagation of electromagnetic waves in vacuum  \emph{Phys. Lett.
A} {\bf 336} 89--96


\bibitem{Peebles}
Peebles P J E 1993 {\em Principles of Physical Cosmology\/}
(Princeton, NJ, USA: Princeton University Press)



\bibitem{Hoff}
Hoffman J D 1992 {\em Numerical Methods for Engineers and
Scientists \/} (New York, USA: McGraw--Hill)

\bibitem{mk2004}
Myung Y S and Kim N J 2004 Difference between anti--de Sitter and
de Sitter spaces: wave equation approach \emph{Class. Quantum
Grav.} {\bf 21} 63--81

\bibitem{rs1}
Randall L and Sundrum R 1999
A large mass hierarchy from a small extra dimension
\emph{Phys. Rev. Lett.} {\bf 83}, 3370--3373

\bibitem{rs2}
Randall L and Sundrum R 1999
An alternative to compactification
\emph{Phys. Rev. Lett.} {\bf 83}, 4690--4693

\bibitem{add}
Arkani--Hamed N, Dimopoulos S and Dvali G 1998 The hierarchy
problem and new dimensions at a millimeter \emph{Phys. Lett. B}
{\bf 429}, 263--272

\bibitem{bajc}
Bajc B and Gabadadze G 2000
Localization of matter and cosmological constant on a brane in anti de Sitter space
\emph{Phys. Lett. B} {\bf 474}, 282--291

\bibitem{pen1}
Pendry J B and Ramakrishna S A 2003
Focusing light using negative refraction
\emph{J. Phys.: Condens. Matter} {\bf 15}, 6345--6364


\end{thebibliography}
\end{document}